\documentclass[11pt]{article}
\usepackage[]{acl}
\usepackage{times}
\usepackage{latexsym}
\usepackage{amsmath}
\usepackage{amssymb}
\usepackage{booktabs}
\usepackage{graphicx}
\usepackage{multirow}
\usepackage{xcolor}
\usepackage{subcaption}
\usepackage{algorithm}
\usepackage{algorithmic}
\usepackage{float}

\newcommand{\lasthop}{\textsc{LastHop}}
\newcommand{\fullsup}{\textsc{FullSup}}
\newcommand{\graphwin}{\textsc{GraphWin}}
\newcommand{\graphloss}{\textsc{GraphLoss}}
\newcommand{\system}{\textsc{PhaseGraph}}

\title{Calibrated Fusion for Heterogeneous Graph-Vector Retrieval in Multi-Hop QA}

\author{Andre Bacellar \\
  \texttt{andremi@gmail.com}}

\begin{document}
\maketitle

\begin{abstract}
Graph-augmented retrieval combines dense similarity with graph-based relevance signals such as Personalized PageRank (PPR), but these scores have different distributions and are not directly comparable. We study this as a \textbf{score calibration problem} for heterogeneous retrieval fusion in multi-hop question answering. Our method, \system{}, maps vector and graph scores to a common unit-free scale using percentile-rank normalization (PIT) before fusion, enabling stable combination without discarding magnitude information.

Across MuSiQue and 2WikiMultiHopQA, calibrated fusion improves held-out last-hop retrieval on HippoRAG2-style benchmarks: \lasthop{}@10 increases from 69.1\% to 71.0\% on MuSiQue (15W/5L, $p{=}0.041$, $n{=}514$) and \lasthop{}@5 from 51.7\% to 53.6\% on 2WikiMultiHopQA (11W/2L, $p{=}0.023$, $n{=}491$), both on independent held-out test splits. Against the official HippoRAG~2 pipeline on 2WikiMultiHopQA, calibrated fusion is ahead at \lasthop{}@10 ($+6.3$pp, $p{<}10^{-3}$) and behind at \lasthop{}@5 ($-8.4$pp), a cutoff-dependent cross-over we report in full. A theory-driven ablation shows that percentile-based calibration is directionally more robust than min-max normalization on both tune and test splits (1W/6L, $p{=}0.125$), while Boltzmann weighting performs comparably to linear fusion after calibration (0W/3L, $p{=}0.25$). These results suggest that \textbf{score commensuration is a robust design choice}, and the exact post-calibration operator appears to matter less on these benchmarks.
\end{abstract}

\section{Introduction}
\label{sec:intro}

Multi-hop question answering requires retrieving multiple evidence passages that form a reasoning chain, where later-hop passages are often weakly aligned with the original query. Dense retrieval is strong at lexical and semantic matching, but misses bridge passages whose relevance is mediated by graph structure rather than direct textual similarity. Graph-based retrieval offers a complementary signal through connectivity and diffusion, for example via Personalized PageRank (PPR) over an entity graph \citep{gutierrez2024hipporag, gutierrez2025hipporag2}.

A practical difficulty is that dense and graph retrieval scores are \textbf{heterogeneous}: cosine similarities cluster in a narrow Gaussian ($\mu \approx 0.09$, $\sigma \approx 0.02$), while PPR scores follow a power-law (most values $\sim 0.001$, rare peaks $\sim 0.3$). Na\"ive score fusion is poorly behaved under this mismatch, while rank-only fusion (RRF) avoids calibration at the cost of discarding magnitude information. We study graph-vector retrieval fusion through the lens of \textbf{score commensuration}.

We present \system{}, a calibrated fusion method for heterogeneous graph-vector retrieval. The method first maps each retriever's scores to a common unit-free scale using percentile-rank normalization (the probability integral transform, PIT), then combines the calibrated signals with a fusion rule. The central hypothesis is that \emph{calibration is the first-order step}: once scores are made commensurable, the exact post-calibration operator appears to matter less than in uncalibrated fusion.

We evaluate this on multi-hop benchmarks under both an earlier pipeline and a stronger HippoRAG2-style setup. Our contributions:

\begin{itemize}
\item \textbf{Calibrated heterogeneous-score fusion} (Sections~\ref{sec:hipporag2},\,\ref{sec:theory-ablation}): We identify heterogeneous-score calibration as the central problem in graph-vector retrieval fusion, and propose PIT-based normalization. Held-out ablation is consistent with PIT being directionally more robust than min-max (1W/6L, $p{=}0.125$); the fusion operator (Boltzmann vs.\ linear) is empirically inconclusive (0W/3L, $p{=}0.25$).

\item \textbf{Held-out confirmation on two benchmarks} (Section~\ref{sec:hipporag2}): Under a strong HippoRAG2-style pipeline (Llama 3.3 70B FP8, NV-Embed-v2 4096d), \system{} improves last-hop retrieval on held-out test splits of MuSiQue (15W/5L, $p{=}0.041$) and 2WikiMultiHopQA (11W/2L, $p{=}0.023$). True RRF also confirms on MuSiQue (25W/11L, $p{=}0.029$), where the two fusions do not separate from each other, but not on 2Wiki (8W/2L, $p{=}0.109$).

\item \textbf{Scale analysis and secondary evidence} (Section~\ref{sec:secondary}): Pool explosion and entity fragmentation explain why uncalibrated fusion fails at corpus scale. Pool capping plus synonym linking restore zero-loss behavior (8W/0L, $p{=}0.008$). We also confirm empirically that cross-encoder reranking destroys multi-hop retrieval (Appendix~\ref{app:stats}), while Ising graph reranking adds marginal exploratory benefit on the curated subset only.
\end{itemize}

\section{Related Work}
\label{sec:related}

\paragraph{Score Fusion in IR.}
Reciprocal Rank Fusion \citep[RRF;][]{cormack2009reciprocal} discards scores entirely, using only ranks. CombSUM and CombMNZ \citep{fox1993combination} assume commensurable scores and apply additive or multiplicative combination. \citet{montague2001condorcet} show that normalization matters more than the fusion algorithm, but leave the normalization choice unresolved. Our work tests this insight in the graph-vector setting, finding that percentile-rank normalization is a robust choice in this setting.

BoltzRank \citep{volkovs2009boltzrank} applies Boltzmann distributions to permutations for learning-to-rank, but not to score fusion. Plackett-Luce models \citep{hunter2004mm} estimate item strengths from ranked lists via maximum likelihood, but assume homogeneous rankers. In our exploratory sweep, Plackett-Luce underperforms Boltzmann fusion when ranker score distributions differ fundamentally (Appendix~\ref{app:ablation-bars}).

\paragraph{Graph-Augmented RAG.}
HippoRAG \citep{gutierrez2024hipporag} and HippoRAG~2 \citep{gutierrez2025hipporag2} use Personalized PageRank for passage-level retrieval, achieving R@5$=$74.7\% on MuSiQue. PropRAG \citep{wang2025proprag} extends this with beam search over the knowledge graph (R@5$=$77.3\%, QA F1$=$78.3\%). More recent energy-based approaches \citep{yu2025graphflow} extend graph-RAG further, but require substantial additional LLM calls per query.

Our approach differs in that we fuse \emph{existing} vector and graph retrieval scores rather than designing new traversal algorithms. In principle, this makes the fusion step applicable to any graph-RAG system that produces two score lists, though we evaluate only our own pipeline.

\paragraph{Multi-Hop Retrieval and Reranking.}
Standard cross-encoders evaluate passages independently and can demote bridge evidence in multi-hop QA. We confirm this empirically: cross-encoder reranking causes catastrophic last-hop failure on our benchmarks (Appendix~\ref{app:stats}).

Our entity extraction follows a similarly minimal approach, avoiding full relation extraction to reduce construction cost.

\section{Calibrated Fusion for Heterogeneous Retrieval}
\label{sec:method}

\subsection{Problem Setting}

Given a query $q$, a vector retriever returns $\mathcal{R}_v = \{(d_i, s^v_i)\}_{i=1}^{N_v}$ (typically $N_v = 10$), and a graph retriever returns $\mathcal{R}_g = \{(d_j, s^g_j)\}_{j=1}^{N_g}$ (typically $N_g \gg N_v$, as PPR returns all reachable nodes). The challenge: $s^v$ and $s^g$ are on incomparable scales and follow different distributions.

\subsection{Percentile-Rank Normalization}

For each system $k \in \{v, g\}$, we compute the percentile rank of each document within its own score list:
\begin{equation}
\hat{p}_i^k = \frac{|\{j : s_j^k \leq s_i^k\}|}{N_k}
\end{equation}
This is simply the empirical CDF, mapping scores to $[0, 1]$ uniform (equivalently, the 1D optimal transport map to $\mathcal{U}[0,1]$ \citep{brenier1991polar}). This preserves within-system ordering while making cross-system values commensurable.

\subsection{Boltzmann Energy and Temperature}

We convert percentile ranks to energies:
\begin{equation}
E_i^k = -\ln(\hat{p}_i^k + \epsilon)
\end{equation}
where $\epsilon = 10^{-6}$ prevents singularity at $\hat{p} = 0$. High-ranked documents have low energy (high percentile $\rightarrow$ small $E$).

We set temperature as a fraction of mean energy:
\begin{equation}
T_k = \frac{\bar{E}_k}{2} = \frac{1}{2N_k}\sum_{i=1}^{N_k} E_i^k
\end{equation}
The factor of $\frac{1}{2}$ is a heuristic that produces moderate sharpening; we did not tune this value. The effect is that Boltzmann probabilities are neither too peaked nor too flat, adapting to the spread of each system's scores.

\subsection{Weighted Boltzmann Fusion}

Boltzmann probabilities within each system:
\begin{equation}
P_i^k = \frac{\exp(-E_i^k / T_k)}{Z_k}, \quad Z_k = \sum_j \exp(-E_j^k / T_k)
\end{equation}

Final fusion with mixing parameter $\alpha$ and consensus boost $\beta$:
\begin{equation}
\text{score}(d_i) = \alpha \cdot P_i^v + (1-\alpha) \cdot P_i^g + \beta \cdot \mathbb{1}[d_i \in \mathcal{R}_v \cap \mathcal{R}_g]
\label{eq:fusion}
\end{equation}
where $\alpha$ controls vector-graph weighting and $\beta$ boosts documents found by both systems (CombMNZ-inspired). Note that $\beta$ operates on a different scale than the probability terms $P_i^k$; in practice, $\beta \in \{0.5, 1.0, 1.5\}$ acts as a discrete bonus for consensus documents. The appropriate $\alpha$ depends on corpus size and graph density; empirical calibration is discussed in Section~\ref{sec:secondary}.

The top-$K$ documents by fused score form the final retrieval set.

\subsection{Comparison to RRF}

Reciprocal Rank Fusion computes $\text{score}(d) = \sum_k \frac{1}{k_0 + \text{rank}_k(d)}$, using only ranks. Our method additionally uses the \emph{shape} of the score distribution through Boltzmann weighting: a document ranked 5th out of 10 similar-scoring documents receives different weight than one ranked 5th with a sharp score drop-off. This temperature-mediated confidence is one advantage over RRF.

\section{Experimental Setup}
\label{sec:setup}

\subsection{Primary Benchmarks}

Primary evaluations use two benchmarks under the primary HippoRAG2 pipeline described below:
\begin{itemize}
\item \textbf{MuSiQue} \citep{trivedi2022musique}: 11{,}654 passages, MD5-split into 486-query tune / 514-query held-out test ($K{=}10$). Each query has 2--4 supporting passages forming a reasoning chain; the final hop is hardest to retrieve. \lasthop{} on MuSiQue scores the passage that supports the terminal step of the annotated decomposition, which is the answer-bearing passage for every query.\footnote{Versions 1--3 of this paper scored the last supporting paragraph in MuSiQue's paragraph order, which MuSiQue shuffles; that passage is the terminal-hop passage for only 45\% of test queries. The MuSiQue \lasthop{} figures in Section~\ref{sec:hipporag2} are recomputed against the terminal passage from a fresh-embedding re-run of the same pipeline, which under the old rule reproduces the published figures within one query (vector-only 384 vs.\ 386 hits, \system{} 393 vs.\ 393, 9W/0L vs.\ 8W/1L; RRF 393 vs.\ 395). The secondary MuSiQue analyses (Section~\ref{sec:secondary}) keep the legacy rule, the last supporting paragraph in paragraph order, and should be read as scoring a gold passage rather than the terminal hop.} On 2Wiki, \lasthop{} scores the last listed supporting fact, which is the answer-bearing passage for compositional and inference questions (253 of 254 test queries of those types) and one of the two compared entities for comparison questions.
\item \textbf{2WikiMultiHopQA} \citep{ho2020constructing}: 6{,}343 passages (isolated database), 509-query tune / 491-query held-out test ($K{=}5$).
\end{itemize}

Secondary analyses use a legacy pipeline on MuSiQue full-corpus (500 queries, 6{,}841 passages) and a curated 66-query hard-slice; details in Section~\ref{sec:secondary}.

\subsection{System Architecture}

The core components are shared across both pipelines: a vector store (PostgreSQL + pgvector), a knowledge graph (Neo4j with LLM-extracted entities), Personalized PageRank for graph retrieval, and a pluggable fusion layer. Two pipeline variants appear in this paper:

\begin{itemize}
\item \textbf{HippoRAG2 pipeline} (Sections~\ref{sec:hipporag2},\,\ref{sec:theory-ablation}): NV-Embed-v2 \citep{lee2024nv} (4096d) for dense retrieval; Llama 3.3 70B FP8 for entity extraction. This is the primary benchmark pipeline.
\item \textbf{Legacy pipeline} (Section~\ref{sec:secondary}): text-embedding-3-small (1536d) for dense retrieval; GPT-4-class LLM for entity extraction. Used for exploratory scaling analyses and the curated hard-slice. Results are not directly comparable to the HippoRAG2 pipeline.
\end{itemize}

\subsection{Metrics}

\begin{itemize}
\item \textbf{\lasthop{}@$K$}: Last-hop passage found in top-$K$ (hardest hop, most distant from query)
\item \textbf{\fullsup{}@$K$}: All supporting passages found in top-$K$ (complete reasoning chain)
\item \textbf{\graphwin{}}: Fused retrieval found last hop when vector alone did not
\item \textbf{\graphloss{}}: Fused retrieval lost a last hop that vector alone found
\end{itemize}

\graphwin{} and \graphloss{} provide a Condorcet-style analysis: a fusion method is safe if \graphloss{}$= 0$ and effective if \graphwin{}$> 0$.

\subsection{Baselines and Configurations}

Primary comparisons: vector-only, true RRF \citep{cormack2009reciprocal}, and our calibrated \textbf{Boltzmann fusion} (Section~\ref{sec:method}). Additional fusion strategies (log-linear, power mean, Tsallis, Gumbel copula, Plackett-Luce, and five others) and Ising reranking are reported as exploratory analyses in Section~\ref{sec:secondary} and Appendix~\ref{app:strategies}.

\section{Primary Results: Held-Out Multi-Hop Benchmarks}
\label{sec:hipporag2}

We evaluate \system{} on the HippoRAG2 benchmark setup \citep{gutierrez2025hipporag2} using the primary pipeline (NV-Embed-v2, Llama 3.3 70B FP8) described in Section~\ref{sec:setup}.

\subsection{Setup and Results}

MuSiQue: 11{,}654 passages, 514-query held-out test ($K{=}10$), tune on 486. 2Wiki: 6{,}343 passages, 491-query test ($K{=}5$), tune on 509; isolated database (no MuSiQue contamination). All embeddings via NV-Embed-v2 (4096d); extraction via Llama 3.3 70B FP8.

\begin{table}[t]
\centering
\small
\caption{HippoRAG2 benchmark held-out test results. W/L vs.\ vector only on \lasthop{}@10 for MuSiQue ($n{=}514$, terminal-hop passage) and \lasthop{}@5 for 2Wiki ($n{=}491$). Configurations fixed on the tune split; MuSiQue test figures recomputed under the terminal-hop metric from a fresh-embedding re-run (footnote in Section~\ref{sec:setup}).}
\label{tab:hipporag2}
\resizebox{\columnwidth}{!}{%
\begin{tabular}{@{}llrrrr@{}}
\toprule
\textbf{Bench.} & \textbf{Strategy} & \textbf{\lasthop} & \textbf{W} & \textbf{L} & \textbf{$p$} \\
\midrule
\multirow{3}{*}{MuSiQue (@10)} & Vector only & 69.1 & -- & -- & -- \\
 & True RRF & 71.8 & 25 & 11 & .029 \\
 & \system{} (.7, dk30) & \textbf{71.0} & \textbf{15} & \textbf{5} & \textbf{.041} \\
\midrule
\multirow{3}{*}{2Wiki (@5)} & Vector only & 51.7 & -- & -- & -- \\
 & True RRF & 53.0 & 8 & 2 & .109 \\
 & \system{} (.4, dk20) & \textbf{53.6} & \textbf{11} & \textbf{2} & \textbf{.023} \\
\bottomrule
\end{tabular}%
}
\end{table}

Table~\ref{tab:hipporag2} shows that \system{} outperforms vector-only on both benchmarks at held-out test, confirming the calibrated fusion approach generalizes beyond the older extraction pipeline. On MuSiQue both fusions confirm against vector-only: RRF at 25W/11L ($p{=}0.029$) and \system{} at 15W/5L ($p{=}0.041$); paired against each other they do not separate (7W/11L, $p{=}0.48$). \system{}'s MuSiQue wins concentrate on 2-hop questions (10W/0L); on 3- and 4-hop questions it is 5W/5L. On 2Wiki, RRF again does not confirm (8W/2L, $p{=}0.109$) but \system{} does (11W/2L, $p{=}0.023$).

\begin{figure}[t]
\centering
\includegraphics[width=\columnwidth]{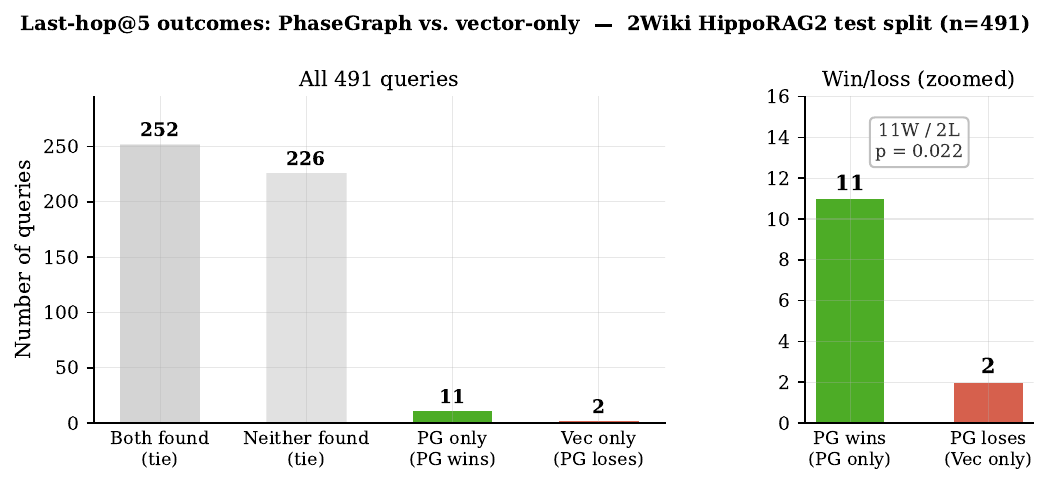}
\caption{Per-query last-hop outcomes: \system{} vs.\ vector-only on 2Wiki HippoRAG2 test split (n$=$491). \emph{Left}: all 491 queries; \emph{right}: zoomed win/loss. The 11:2 asymmetry gives $p{=}0.022$ \citep{mcnemar1947}.}
\label{fig:perquery}
\end{figure}

Figure~\ref{fig:perquery} shows the per-query last-hop outcome breakdown for 2Wiki: 252 queries where both systems find the last hop, 226 where neither does, and the critical asymmetry of 11 \system{}-only wins vs.\ 2 vector-only wins.

\paragraph{Comparison to published baselines.} HippoRAG~2 reports R@5$=$74.7\% on MuSiQue; PropRAG reports R@5$=$77.3\% (QA F1: 78.3\%). Note that R@5 (any gold passage in top-5) and \lasthop{}@5 (final hop in top-5) are different metrics---\lasthop{} is the harder, more diagnostic one for multi-hop reasoning and is not directly comparable. Our vector-only \lasthop{}@10 on MuSiQue is 69.1\%, reflecting the strong NV-Embed-v2 baseline; \system{} adds $+$1.9pp where pool explosion and entity noise are the binding constraint.

\paragraph{Head-to-head against the official HippoRAG~2 pipeline.} Published R@5 figures are not directly comparable to \lasthop{}, so we ran the official HippoRAG~2 implementation on the same 2Wiki held-out split ($n{=}491$) with the same NV-Embed-v2 encoder and scored both systems on \lasthop{}. The outcome depends on the cutoff. At \lasthop{}@10, \system{} reaches 76.99\% against HippoRAG~2's 70.67\% ($+6.32$pp, paired 52W/21L, $p{=}3.7\times10^{-4}$). At \lasthop{}@5, HippoRAG~2 is ahead: 62.32\% against our 53.97\% ($-8.35$pp, 2W/43L, $p{=}5.9\times10^{-11}$). HippoRAG~2's own vector-only baseline sits at 53.56\% / 61.10\% (@5 / @10). We read this as a cross-over rather than a win: commensurated fusion recovers last-hop evidence that dense retrieval ranks in positions 6--10, while HippoRAG~2's graph step is stronger at pushing it into the top five. Which matters depends on how many passages the downstream reader is given.

We also ran a diagnostic R@5 probe on all 1000 HippoRAG2 benchmark queries (Appendix~\ref{app:r5sweep}). Our vector-only baseline reaches R@5$=$69.8\% overall (70.1\% on the test split), with a pronounced hop-count gradient: 77.9\% on 2-hop queries---already above HippoRAG~2's 74.7\%---but 68.9\% on 3-hop and 46.2\% on 4-hop. Thermo fusion does not improve this strict top-5 metric (0W/0L, $p{=}1.0$), whereas it confirms at $k{=}10$ in the primary evaluation. This contrast is consistent with a slot-competition bottleneck: at $k{=}5$, graph bridge passages displace rather than supplement co-gold passages. Graph evidence is therefore more useful for expanding recall in a wider candidate set---the operating regime studied in the primary evaluation above.

\section{What Matters in Fusion? A Theory-Guided Ablation}
\label{sec:theory-ablation}

We run a theory-driven ablation on the 2Wiki HippoRAG2 setup to isolate the contribution of each design decision. All ablations use the confirmed baseline ($\alpha{=}0.4$, dk$=$20).

\subsection{Score Distributions and the Calibration Motivation}

\begin{figure}[t]
\centering
\includegraphics[width=\columnwidth]{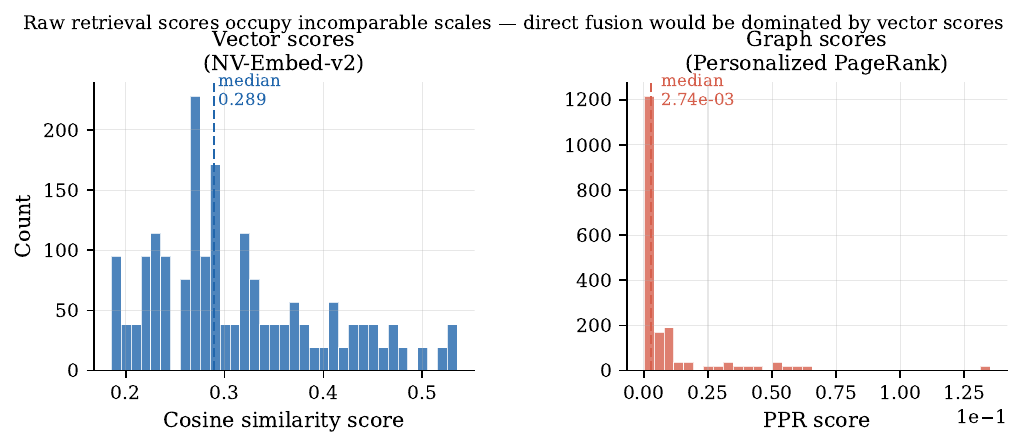}
\caption{Raw score distributions from 100 sample queries (2Wiki HippoRAG2). Vector cosine similarity and PPR scores occupy incomparable scales; direct fusion would be dominated by vector scores.}
\label{fig:score-dist}
\end{figure}

\begin{figure}[t]
\centering
\includegraphics[width=\columnwidth]{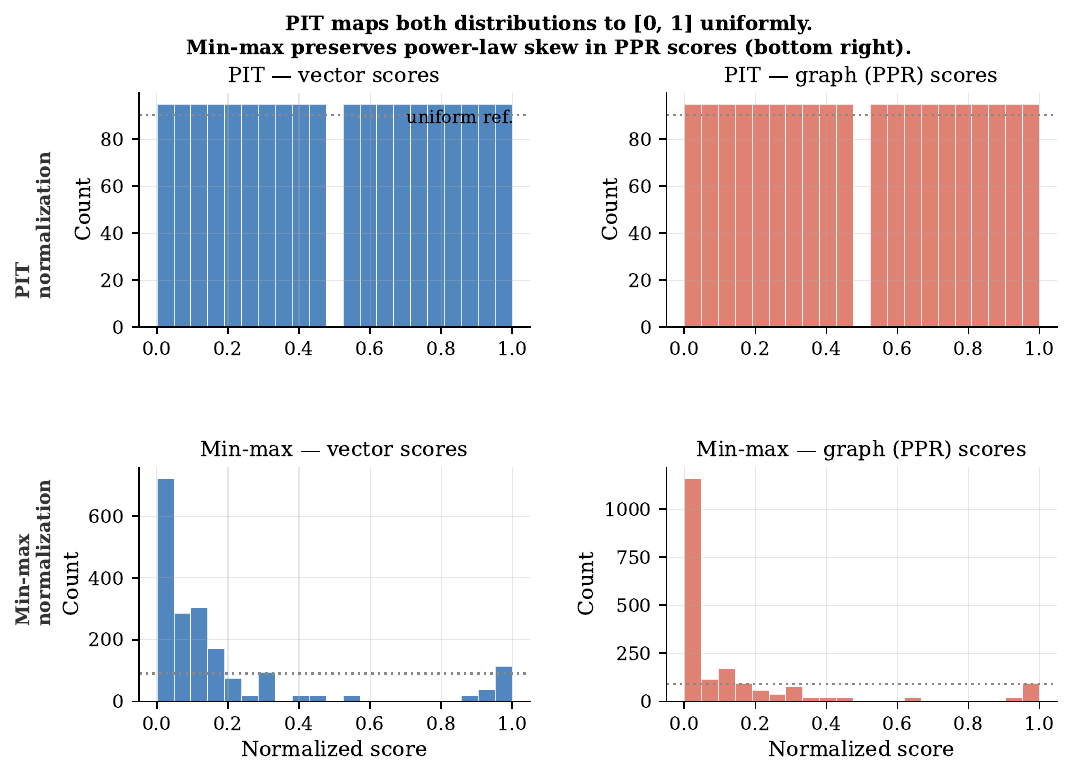}
\caption{Effect of normalization. \emph{Top row}: PIT maps both distributions to approximately uniform $[0,1]$, making them commensurable. \emph{Bottom row}: min-max normalization preserves the power-law spike in PPR scores, producing unequal marginals. Dashed line: expected uniform count per bin.}
\label{fig:norm-effect}
\end{figure}

Figure~\ref{fig:score-dist} shows raw score distributions from 100 sample queries. Vector cosine similarity spans $[0.19, 0.54]$ with median 0.29. PPR scores span $[6\text{e-}5, 0.14]$ with a heavy right tail---effectively 6$\times$ smaller median. Direct weighted combination would be dominated by vector scores regardless of $\alpha$. Figure~\ref{fig:norm-effect} confirms that PIT normalization maps both distributions to approximately uniform $[0,1]$, while min-max normalization preserves the power-law skew in PPR scores.

\subsection{Ablation Results}

We test three axes: (1)~normalization method, (2)~fusion formula, (3)~temperature calibration. Table~\ref{tab:theory-ablation} reports tune-split (n$=$509) and held-out test-split (n$=$491) results.

\begin{table}[t]
\centering
\small
\caption{Theory ablation on 2WikiMultiHopQA HippoRAG2 (tune n$=$509, test n$=$491). W/L vs.\ baseline on \lasthop{}@5. Test split evaluated once (held-out).}
\label{tab:theory-ablation}
\resizebox{\columnwidth}{!}{%
\begin{tabular}{@{}llrrrrr@{}}
\toprule
\multirow{2}{*}{\textbf{Config}} & \multirow{2}{*}{\textbf{Axis}} & \multicolumn{2}{c}{\textbf{Tune}} & \multicolumn{3}{c}{\textbf{Test (held-out)}} \\
\cmidrule(lr){3-4}\cmidrule(lr){5-7}
 & & \textbf{\lasthop} & \textbf{W/L} & \textbf{\lasthop} & \textbf{W/L} & \textbf{$p$} \\
\midrule
Baseline (PIT + Boltzmann) & -- & 52.5 & -- & 53.8 & -- & -- \\
\midrule
Min-max norm & Norm. & 51.7 & 1/5 & 52.7 & 1/6 & .125 \\
Raw/max norm & Norm. & 52.8 & 5/3 & -- & -- & -- \\
\midrule
Linear fusion & Formula & 52.5 & 2/2 & 53.2 & 0/3 & .250 \\
\midrule
Fixed $T{=}0.3$ & Temp. & 52.1 & 2/4 & -- & -- & -- \\
Fixed $T{=}0.6$ & Temp. & 52.1 & 1/3 & -- & -- & -- \\
Fixed $T{=}1.0$ & Temp. & 52.5 & 1/1 & -- & -- & -- \\
\bottomrule
\end{tabular}%
}
\end{table}

\paragraph{Normalization.}
Min-max normalization is directionally worse than PIT on both splits (tune 1W/5L, test 1W/6L $p{=}0.125$), consistent with Figure~\ref{fig:norm-effect}: min-max preserves the PPR power-law skew, producing unequal marginals that make the fusion weight $\alpha$ non-stationary. Raw normalization (score/max) is roughly equivalent to PIT on the tune split (5W/3L), likely because the within-system score distributions are smooth enough that raw/max approximates percentile-rank on this data. We recommend PIT for distributional robustness.

\paragraph{Fusion formula.}
Replacing Boltzmann weighting with linear combination of PIT-normalized scores is directionally worse on held-out test (0W/3L, $p{=}0.25$) but inconclusive---the tune result is tied (2W/2L). This is consistent with the hard-slice finding (Appendix~\ref{app:ablation-bars}) that normalization appears to be the dominant empirical factor on this benchmark. Boltzmann is a principled choice with theoretical motivation (Section~\ref{sec:method}), but it is not the empirically decisive mechanism on these benchmarks.

\paragraph{Temperature calibration.}
Low fixed temperatures ($T{=}0.3$, $T{=}0.6$) are directionally worse than auto-calibration, consistent with the interpretation that low $T$ produces overconfident Boltzmann weights. Fixed $T{=}1.0$ is roughly equivalent to auto-calibration on this data.

\section{Secondary Analyses: Scaling, Isolation, and Exploratory Evidence}
\label{sec:secondary}

\textit{The analyses in this section are exploratory or secondary; primary confirmatory evidence is in Section~\ref{sec:hipporag2}.}

\subsection{Hard-Slice: Exploratory Evidence}

On a curated 66-query subset selected for KG coverage (668 passages), all PIT-normalized strategies reach 26 \lasthop{}@10 wins; Ising reranking adds one more (27W, 80.3\%). Zero losses across 660+ configs reflect low error correlation on this favorable subset---a property that does \emph{not} hold at full corpus. Cross-encoder reranking causes catastrophic failure (78.8\%\,$\to$\,9.1\%), confirming that multi-hop retrieval requires passage-dependency-aware scoring: standard cross-encoders evaluate passages independently and demote bridge evidence. Full statistical details (Wilson CIs, odds ratios) appear in Appendix~\ref{app:stats}; the 13-strategy ablation is shown in Figure~\ref{fig:ablation} (Appendix~\ref{app:sweep}); the hard-slice retrieval table is Table~\ref{tab:main} (Appendix~\ref{app:hardslice}). We also explored a mean-field Ising reranker that propagates query relevance through KG entity connections, adding one win on the hard-slice (27 total). This gain does not transfer to full corpus (all Ising configs net negative at scale); full details in Appendix~\ref{app:ising}.

\subsection{Full-Corpus Scaling}
\label{sec:scaling}

Applying the hard-slice configuration ($\alpha{=}0.3$, no pool cap) to 500 MuSiQue queries with 6{,}841 passages causes graph fusion to \emph{hurt retrieval}:

\begin{table}[t]
\centering
\small
\caption{Full-corpus scaling (500 queries, 6{,}841 passages). W/L vs.\ vector only on \lasthop{}@10.}
\label{tab:scaling}
\begin{tabular}{@{}lrrrrr@{}}
\toprule
\textbf{Strategy} & \textbf{\lasthop} & \textbf{R@5} & \textbf{W} & \textbf{L} & \textbf{net} \\
\midrule
Vector only & 58.1 & 88.0 & -- & -- & -- \\
Thermo .3, uncap. & 51.0 & 78.8 & 36 & 81 & $-$45 \\
True RRF, uncap. & 58.6 & 83.4 & 32 & 39 & $-$7 \\
Thermo .7, dk30 & 60.4 & 88.6 & 16 & 5 & +11 \\
\midrule
\multicolumn{6}{@{}l}{\emph{+ Entity disambig.\ (synonym linking):}} \\
Vector only & 60.1 & 89.4 & -- & -- & -- \\
\textbf{Thermo .7, dk30} & \textbf{62.6} & \textbf{89.8} & \textbf{12} & \textbf{0} & \textbf{+12} \\
\bottomrule
\end{tabular}
\end{table}

The mechanism is \textbf{graph candidate pool explosion}: 69{,}678 entities generate 2{,}000+ PPR candidates per query (vs.\ $\sim$200 on the hard-slice), and percentile-rank normalization over this diffuse pool allows low-quality results to dominate when $\alpha{=}0.3$ assigns 70\% graph weight. Introducing \texttt{divergent\_top\_k} (pool cap) and shifting to $\alpha{=}0.7$ recovers the gain: 16W/5L (Table~\ref{tab:scaling}). The 5 remaining losses trace to disconnected entity aliases (e.g., ``United States'' vs.\ ``US''); embedding-based synonym linking (0.85 cosine threshold, 44{,}781 links) restores zero losses: 12W/0L (Table~\ref{tab:scaling}, bottom). A held-out split (247 queries) confirms: \textbf{8W/0L} ($p{=}0.008$, \citealt{mcnemar1947}). R@5 does not improve significantly (7W/5L, $p{=}0.77$). On HotpotQA (200 queries, 95.5\% vector baseline), both fusion methods produce 0W/0L---graph fusion only helps when vector retrieval is insufficient.

Legacy-pipeline 2WikiMultiHopQA results (500 queries, 3{,}188 passages, legacy pipeline) appear in Appendix~\ref{app:2wiki-isolated}. Held-out confirmation on the HippoRAG2 pipeline is reported in Section~\ref{sec:hipporag2}.

\section{Discussion}
\label{sec:discussion}

\paragraph{Zero losses and scale sensitivity.}
\label{sec:discussion-isolation}
On the hard-slice (selected for KG coverage), zero \lasthop{}@10 losses across 660+ configs reflects low error correlation between vector and graph failures. At full corpus, this breaks (81 losses at $\alpha{=}0.3$) due to pool explosion and entity fragmentation; pool capping plus synonym linking restore it (held-out: 8W/0L, $p{=}0.008$). On 2WikiMultiHopQA (isolated database), a conservative configuration achieves 15W/1L ($p{<}0.001$) while an aggressive one incurs 7 losses for higher recall (24W/7L, $p{=}0.003$)---loss count is a tunable property. Initial 2Wiki evaluations sharing a Neo4j graph with MuSiQue data (89{,}000 entities combined) produced weaker results across all configurations, consistent with PPR traversals being diluted by foreign entities. Isolated-database legacy-pipeline results appear in Appendix~\ref{app:2wiki-isolated}. On MuSiQue-500 (text-embedding-3-small pipeline), isolated thermo is positive on the full set (17W/6L, $p{=}0.035$) but does not confirm on held-out (5W/4L, $p{=}1.0$); on the HippoRAG2 pipeline (NV-Embed-v2), thermo confirms on the test split (15W/5L, $p{=}0.041$, Section~\ref{sec:hipporag2}).

\paragraph{Normalization ablation.}
Two complementary experiments are consistent with score commensuration being an important contributor to the gains in our experiments. On the hard-slice, PIT with simple averaging gives 23W/1L vs.\ 0W/0L without normalization; Boltzmann adds zero wins over averaging. On 2Wiki HippoRAG2 (Section~\ref{sec:theory-ablation}), min-max is directionally worse than PIT on both splits; raw normalization roughly ties PIT. Full results and analysis in Section~\ref{sec:theory-ablation}.

\paragraph{Limitations.}
Primary held-out results (Section~\ref{sec:hipporag2}) are on MuSiQue and 2WikiMultiHopQA; a HotpotQA cross-check shows no effect (0W/0L) due to a strong vector baseline (Section~\ref{sec:scaling}). On 2WikiMultiHopQA (legacy pipeline; Appendix~\ref{app:2wiki-isolated}), score-based fusion requires per-corpus calibration ($\alpha$, pool cap) to transfer; low loss counts require conservative tuning ($\alpha{=}0.5$, dk$=$50: 15W/1L on held-out). The 2Wiki held-out split is 238/262 (not balanced) and 38/500 queries have incomplete graph coverage, biasing all graph methods downward. On MuSiQue-500 with the legacy text-embedding-3-small pipeline, thermo does not confirm on held-out (5W/4L, $p{=}1.0$); the primary HippoRAG2-pipeline result (15W/5L, $p{=}0.041$, Table~\ref{tab:hipporag2}) supersedes this under the stronger NV-Embed-v2 backbone. That MuSiQue result is sensitive to the query-embedding realization: NV-Embed-v2 query vectors produced under a slightly different instruction wrapper (cosine 0.97 to the ones used here) raise the vector-only baseline to 71.0\% and leave \system{} at 71.2\% (7W/6L, $p{=}1.0$), so the MuSiQue margin should be read as fragile; the 2Wiki result reproduces within one query across embedding batches. The 66-query hard-slice is filtered for KG coverage, making it favorable to graph methods by construction. Testing 660+ configs on 66 queries creates degrees of freedom; the full-corpus config was selected on the same 500 queries used to report 12W/0L (exploratory); confirmatory evidence is the held-out split (8W/0L, $p{=}0.008$). Full-corpus gains are small ($+$2.5pp \lasthop{}@10); R@5 is not significant (7W/5L, $p{=}0.77$). Hard-slice and full-corpus scaling analyses (Section~\ref{sec:secondary}) use text-embedding-3-small (1536d); primary held-out results (Section~\ref{sec:hipporag2}) and the theory ablation (Section~\ref{sec:theory-ablation}) use NV-Embed-v2 (4096d). The 0.85 synonym linking threshold was not tuned. The QA evaluation (Table~\ref{tab:e2e}) uses 66 queries with one LLM and no repeated runs. KG construction requires LLM-based entity extraction (14s/paragraph). The database isolation experiment (Section~\ref{sec:discussion-isolation}) is observational---results differ between mixed and isolated databases, providing strong evidence of cross-dataset contamination but not a tightly controlled causal ablation.

\section{Conclusion}

We presented \system{}, a calibrated fusion method for heterogeneous graph-vector retrieval in multi-hop QA. Our main claim is not that a particular thermodynamic operator is uniquely necessary, but that graph and vector retrieval should be fused only \emph{after} their scores are made commensurable. Across held-out evaluations on MuSiQue and 2WikiMultiHopQA under a strong HippoRAG2-style pipeline, calibrated fusion improves last-hop retrieval over vector-only under the reported configurations on both benchmarks. Neither benchmark establishes a paired advantage over rank-based fusion: thermo versus RRF is 7W/11L ($p{=}0.48$) on MuSiQue and 3W/0L ($p{=}0.25$) on 2Wiki.

A theory-guided ablation clarifies the mechanism. Percentile-based calibration is directionally more robust than min-max normalization on both tune and held-out test splits, while Boltzmann weighting is empirically comparable to linear fusion after calibration. This suggests that the main contribution of \system{} is calibrated heterogeneous-score fusion rather than a specific post-calibration operator.

More broadly, our results argue for treating graph-vector retrieval as a calibration problem before it is treated as an architecture problem. In this setting, on our benchmarks, normalization choice appears to matter more than increasingly elaborate fusion rules, and graph quality constraints---entity disambiguation, pool capping, and database isolation---can matter as much as the fusion formula itself.

\bibliography{references}

\appendix

\section{Theory Ablation Bar Chart}
\label{app:ablation-bars}

\begin{figure}[H]
\centering
\includegraphics[width=\columnwidth]{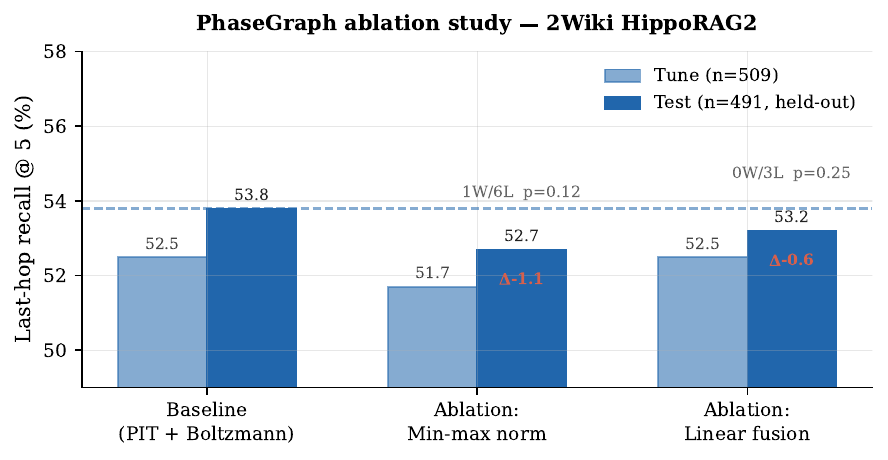}
\caption{Theory ablation: \lasthop{}@5 on 2Wiki HippoRAG2 for baseline vs.\ normalization and fusion ablations. Bars show tune (n$=$509, lighter) and held-out test (n$=$491, solid). Annotations: W/L vs.\ baseline and $\Delta$ \lasthop{} on test bars.}
\label{fig:ablation-theory}
\end{figure}

\section{Hard-Slice Statistical Significance}
\label{app:stats}

\begin{table}[H]
\centering
\small
\caption{\lasthop{}@10 with Wilson 95\% CIs, McNemar \citep{mcnemar1947} $p$-values (vs.\ vector only), and odds ratios. MuSiQue hard-slice, $n{=}66$.}
\label{tab:significance}
\begin{tabular}{@{}lcccc@{}}
\toprule
\textbf{Strategy} & \textbf{Recall} & \textbf{95\% CI} & \textbf{$p$} & \textbf{OR} \\
\midrule
Vector only & 39.4 & {[}28.5, 51.5{]} & --- & --- \\
True RRF & 74.2 & {[}62.6, 83.3{]} & $2.4\text{e-}7$ & 4.43 \\
Thermo & \underline{78.8} & {[}67.5, 87.2{]} & $3.0\text{e-}8$ & \underline{5.71} \\
\textbf{Thermo+Ising} & \textbf{80.3} & {[}69.2, 88.2{]} & $1.5\text{e-}8$ & \textbf{6.27} \\
Thermo+CE & 9.1 & {[}4.2, 18.4{]} & $3.6\text{e-}5$ & 0.15 \\
\bottomrule
\end{tabular}
\end{table}

\section{Fusion Strategy Details}
\label{app:strategies}

\paragraph{True RRF.} $\text{score}(d) = \sum_k \frac{1}{60 + \text{rank}_k(d)}$ \citep{cormack2009reciprocal}.

\paragraph{Log-Linear.} $\text{score}(d) = P_v(d)^\alpha \cdot P_g(d)^{1-\alpha}$, where $P_k$ are Boltzmann probabilities. Equivalent to geometric mean in probability space.

\paragraph{Power Mean.} $\text{score}(d) = (\alpha \cdot \hat{p}_v^p + (1-\alpha) \cdot \hat{p}_g^p)^{1/p}$, using raw percentile ranks. Sub-arithmetic mean ($p < 1$) rewards consensus.

\paragraph{Tsallis $q$-Exponential.} Replaces $\exp(-E/T)$ with $[1 + (1-q) \cdot (-E/T)]_+^{1/(1-q)}$, allowing heavier tails for the graph system ($q > 1$).

\paragraph{Gumbel Copula.} $C_\theta(u, v) = \exp(-((-\ln u)^\theta + (-\ln v)^\theta)^{1/\theta})$ with $\theta$ estimated from overlap documents.

\paragraph{Plackett-Luce MLE.} MM algorithm \citep{hunter2004mm} estimates strengths $\gamma_i$ from each system's ranking, combined via $s_i = \gamma_{i,v}^\alpha \cdot \gamma_{i,g}^{1-\alpha}$.

\paragraph{Quantum Interference.} $P = P_v + P_g + 2\sqrt{P_v P_g}\cos\theta$, with fixed $\theta = 0$ (constructive only). Based on QPRP \citep{zuccon2009quantum}.

\paragraph{OT Alignment.} Optimal transport map from each score distribution to Gaussian target, then additive fusion.

\paragraph{Wasserstein-T.} Temperature modulated by Wasserstein-1 distance between score distributions: $T = T_0(1 + \gamma W_1)$.

\section{Ising Parameter Sweep}
\label{app:ising}

\begin{figure}[H]
\centering
\includegraphics[width=\columnwidth]{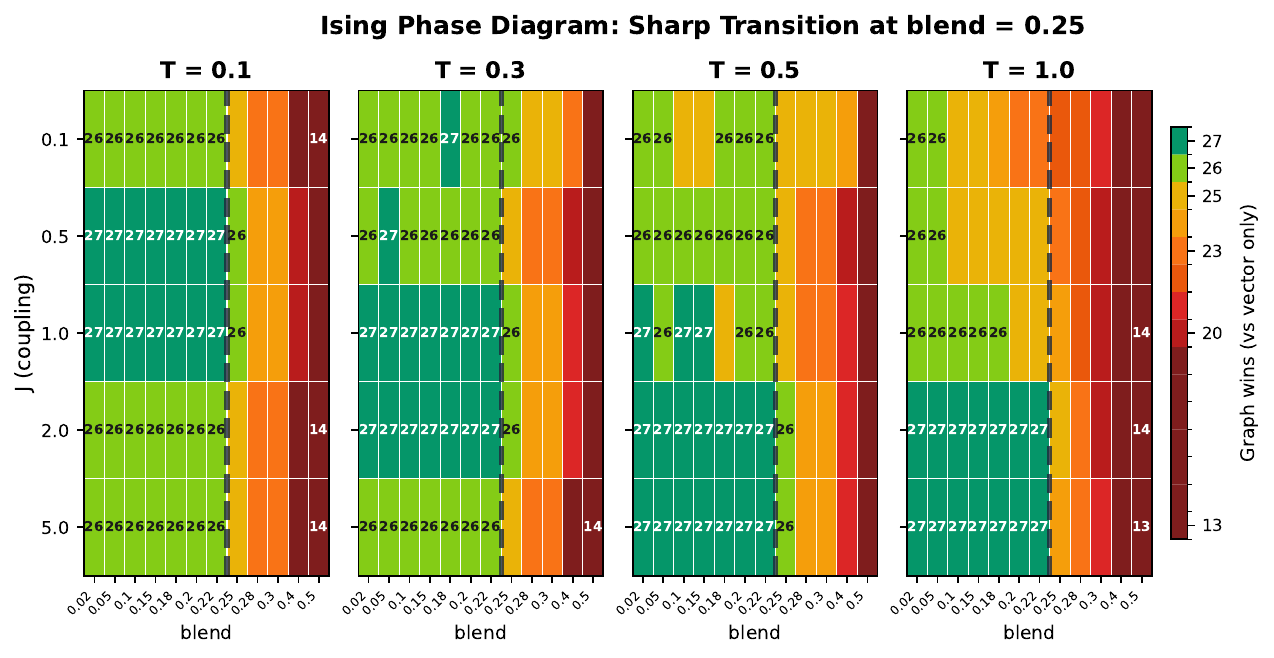}
\caption{Ising parameter sweep. Sharp threshold at blend$=$0.25 across all $(J, T)$ pairs. Green: 27 wins, gold: 26, red gradient: $\leq$25.}
\label{fig:phase}
\end{figure}

The blend parameter governs a regime change. Below 0.25, graph structure constructively supplements fusion (mean 26.3 wins). Above, coupling increasingly disrupts ordering. The $J/T$ ratio governs within-phase behavior: high $T$ requires strong $J$ to propagate relevance.

\section{Full Sweep Results}
\label{app:sweep}

\begin{table}[H]
\centering
\small
\caption{Configuration sweep summary across three tiers. \textbf{Bold}: best per column.}
\label{tab:sweep}
\begin{tabular}{@{}lccc@{}}
\toprule
\textbf{Category} & \textbf{\#Cfg} & \textbf{Best W} & \textbf{Max \lasthop} \\
\midrule
Tier 1 (fusion) & 102 & 26 & 78.8 \\
Tier 2 (reranking) & \textbf{277} & \textbf{27} & \textbf{80.3} \\
Tier 3 (research) & 288 & \textbf{27} & \textbf{80.3} \\
\midrule
Total & 667 & \textbf{27} & \textbf{80.3} \\
\bottomrule
\end{tabular}
\end{table}

\section{End-to-End QA Results}
\label{app:e2e}

Table~\ref{tab:e2e} shows that retrieval improvements translate to answer quality on the curated hard-slice. Fused retrieval with 10 passages significantly outperforms context stuffing with 20 passages (including all gold; $\Delta{=}+0.089$ F1, $p{=}0.028$); Thermo+Ising outperforms stuffing by $+$0.124 ($p{=}0.002$).

The gap between vector-only (.135 F1) and Thermo (.271 F1) is larger than the retrieval gap alone would predict, consistent with the last-hop passage being disproportionately important for answer generation: retrieving the final bridge passage enables the LLM to complete the reasoning chain, while missing it degrades the answer even when the other hops are present. Context stuffing (.183 F1) partially closes this gap by including all gold passages, but the noise from 20 passages --- many irrelevant --- limits LLM accuracy.

\textbf{Caveats.} All results use a single LLM (Claude Haiku 4.5) with no repeated runs, on the 66-query curated slice selected for KG coverage. F1 and EM should be treated as exploratory indicators, not benchmark results. The confidence intervals are wide; the ordering is consistent but the absolute values are pipeline-specific.

\begin{table}[H]
\centering
\small
\caption{End-to-end QA on hard-slice (Claude Haiku 4.5, 66 queries). 95\% bootstrap CIs ($n{=}10{,}000$).}
\label{tab:e2e}
\begin{tabular}{@{}lcc@{}}
\toprule
\textbf{Config} & \textbf{F1 [95\% CI]} & \textbf{EM} \\
\midrule
Vector only & .135 [.075, .203] & 6.1\% \\
Context stuff. & .183 [.099, .275] & 15.2\% \\
Thermo & \underline{.271} [.181, .369] & \underline{18.2\%} \\
\textbf{Thermo+Ising} & \textbf{.307} [.213, .406] & \textbf{21.2\%} \\
\bottomrule
\end{tabular}
\end{table}

\section{Isolated 2Wiki Legacy-Pipeline Results}
\label{app:2wiki-isolated}

\begin{table}[H]
\centering
\small
\caption{2WikiMultiHopQA held-out test results (262 queries, isolated database, top-10). W/L vs.\ vector only on \lasthop{}@10. $^\dagger$one query failed (n$=$261). Thermo configs: tune winner ($\alpha{=}0.4$, dk20), best recall ($\alpha{=}0.4$, dk50), safest ($\alpha{=}0.5$, dk50).}
\label{tab:2wiki}
\begin{tabular}{@{}lrrrr@{}}
\toprule
\textbf{Strategy} & \textbf{\lasthop} & \textbf{W} & \textbf{L} & \textbf{$p$} \\
\midrule
Vector only & 37.8 & -- & -- & -- \\
True RRF$^\dagger$ & 41.2 & 15 & 6 & .078 \\
\midrule
Thermo (tune winner) & 43.5 & 26 & 11 & .020 \\
Thermo (best recall) & \textbf{44.3} & \textbf{24} & 7 & \textbf{.003} \\
Thermo (safest) & 43.1 & 15 & \textbf{1} & $<$.001 \\
\bottomrule
\end{tabular}
\end{table}

\section{Hard-Slice Retrieval Table and Ablation}
\label{app:hardslice}

\begin{table}[H]
\centering
\small
\caption{Retrieval on MuSiQue hard-slice (66 queries, top-10). \textbf{Bold}: best; \underline{underline}: tied at PIT-fusion ceiling (78.8\%). $^{***}$\!: $p < 0.001$ vs.\ vector only \citep{mcnemar1947}.}
\label{tab:main}
\begin{tabular}{@{}lcccc@{}}
\toprule
\textbf{Strategy} & \textbf{\lasthop} & \textbf{\fullsup} & \textbf{W} & \textbf{L} \\
\midrule
Vector only & 39.4 & 6.1 & -- & -- \\
Legacy RRF & 39.4 & 6.1 & 0 & 0 \\
True RRF$^{***}$ & 74.2 & 48.5 & 23 & 0 \\
\midrule
Thermo$^{***}$ & \underline{78.8} & \underline{57.6} & 26 & 0 \\
Log-linear$^{***}$ & \underline{78.8} & \underline{57.6} & 26 & 0 \\
Power mean$^{***}$ & \underline{78.8} & \underline{57.6} & 26 & 0 \\
Tsallis$^{***}$ & \underline{78.8} & \underline{57.6} & 26 & 0 \\
Gumbel copula & 77.3 & 56.1 & 25 & 0 \\
Plackett-Luce & 77.3 & 56.1 & 25 & 0 \\
\midrule
\textbf{Thermo+Ising}$^{***}$ & \textbf{80.3} & \textbf{62.1} & \textbf{27} & \textbf{0} \\
\midrule
Thermo+CE$^{***}$ & 9.1 & 1.5 & 2 & 22 \\
\bottomrule
\end{tabular}
\end{table}

\begin{figure}[H]
\centering
\includegraphics[width=\columnwidth]{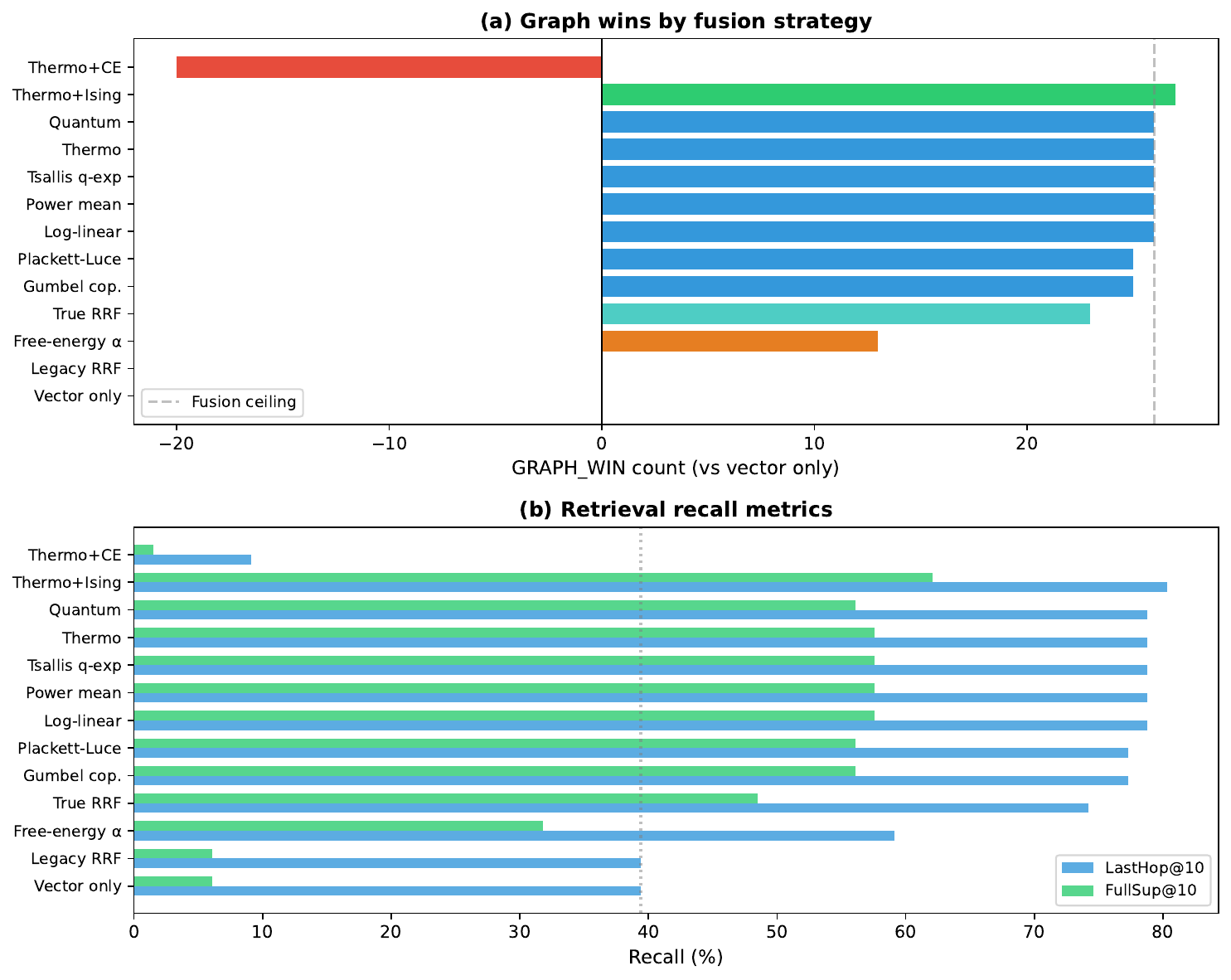}
\caption{Fusion strategy ablation on MuSiQue hard-slice ($n{=}66$, \lasthop{}@10 vs.\ vector-only). Four PIT-normalized strategies tie at the 26-win ceiling; only Ising coupled reranking breaks through to 27.}
\label{fig:ablation}
\end{figure}

\section{MuSiQue R@5 Diagnostic Sweep}
\label{app:r5sweep}

Table~\ref{tab:r5sweep} reports a full R@5 comparison on all 1000 MuSiQue HippoRAG2 queries (486 tune / 514 test), using NV-Embed-v2 embeddings computed via the MDMA retrieval API.
Published baselines use their own embedding pipelines and are shown for orientation only.
The hop-count breakdown reveals the source of the gap: our system matches or exceeds HippoRAG~2 on 2-hop queries (77.9\% vs.\ 74.7\%) but falls behind on 3-hop and especially 4-hop queries.

\begin{table}[H]
\centering
\small
\caption{MuSiQue R@5 on all 1000 HippoRAG2 benchmark queries (486 tune / 514 test). Thermo: $\alpha{=}0.7$, dk$=$30. Published baselines not directly comparable (different pipelines).}
\label{tab:r5sweep}
\resizebox{\columnwidth}{!}{%
\begin{tabular}{@{}lrrr|rrr@{}}
\toprule
\textbf{Method} & \textbf{all} & \textbf{tune} & \textbf{test} & \textbf{2-hop} & \textbf{3-hop} & \textbf{4-hop} \\
\midrule
HippoRAG~2 (pub.)  & 74.7 & -- & -- & -- & -- & -- \\
PropRAG (pub.)     & 77.3 & -- & -- & -- & -- & -- \\
\midrule
Vector only        & \textbf{69.8} & 69.5 & 70.1 & \textbf{77.9} & 68.9 & 46.2 \\
\system{} (thermo) & 69.8 & 69.6 & 70.0 & 77.9 & 69.0 & 46.2 \\
\bottomrule
\end{tabular}%
}
\end{table}

\noindent The any@5 rate is 98.5\%---gold passages are individually retrievable---while full@5 (all gold passages in top-5) is only 37.6\%, consistent with a slot-competition bottleneck that worsens with hop count.
Thermo fusion does not improve R@5 (0W/0L, $p{=}1.0$ on test), in contrast to its confirmed benefit at $k{=}10$ (Section~\ref{sec:hipporag2}).
This is expected: graph bridge passages require additional retrieval slots to avoid displacing co-gold passages within the strict top-5 budget.

\end{document}